\documentclass[12pt]{article}

\def\be{\begin{equation}}
\def\ee{\end{equation}}
\def\bea{\begin{eqnarray}}
\def\eea{\end{eqnarray}}

\begin{document}

\noindent {\small USC-98/HEP-B4\hfill \hfill hep-th/9809034}\newline
{\small \hfill }

{\vskip 0.7cm}

\begin{center}
{\Large TWO-TIME PHYSICS \footnote{{\Large Lecture delivered at the XXII
International Colloquium for Group Theoretical Methods in Physics, Hobart,
Tasmania, Australia, July 1998.}} }\\[0pt]

{\vskip 0.5cm}

{ITZHAK BARS}

{\vskip 0.2cm}

{Department of Physics and Astronomy, University of Southern California}

{\ Los Angeles, CA 90089-0484, USA}

{\vskip 0.5cm}

{\ ABSTRACT}
\end{center}

\noindent {\small We give an overview of the correspondance between
one-time-physics and two-time-physics. This is characterized by the presence
of an SO$\left( d,2\right) $ symmetry and an Sp$\left( 2\right) $ duality
among diverse one-time-physics systems all of which can be lifted to the
same more symmetric two-time-physics system by the addition of gauge degrees
of freedom. We provide several explicit examples of physical systems that
support this correspondance. The example of a particle moving in $
AdS_{D}\times S^{n}$, with SO${(D+n,2)}$ symmetry which is larger than the
popularly known symmetry SO$\left( D-1,2\right) \times $ SO$\left(
n+1\right) $ for this case, should be of special current interest in view of
the proposed AdS-CFT duality.}

\bigskip

\section{Introduction and summary}

One of the conceptual advances around and after Strings-95 is to understand
how to go beyond perturbative String Theory as part of an as yet poorly
understood non-perturbative theory called M-theory. Duality symmetries in
M-theory connect different limits of M-theory including strings in various
backgrounds, p-branes, supergravity and Super Yang-Mills theories. The
supersymmetry structure of the theory has provided global hints of higher
dimensions, including two timelike dimensions \cite{ibtokyo}. Hints of
two-times have been noted from various points of view \cite{ibtokyo}{-} \cite%
{sezrudy}.

If two-time physics is real physics it should be possible to formulate
ordinary physical systems in the language of two time physics {\it without
any ghosts}. In this lecture we will describe some important steps in that
direction. Some technical details and specific examples, including spinning
systems, have appeared in several recent papers \cite{dualconf}{-}\cite%
{dualsusy}. Here we will provide an overview as well as a few new examples
that connect one-time-physics to two-time-physics.

The formalism is a simple Sp$(2,R)$ gauge theory for particles $X^M\left(
\tau \right) $ (zero-branes) which arises from a basic idea as follows. Sp$
(2,R)$ is the automorphism symmetry of the quantum relations $[x,p]=i$ and
treats $(x,p)$ as a doublet. The idea is to turn this global symmetry of
Quantum Mechanics into a local symmetry of a theory. The 3-parameter local
symmetry Sp$(2,R)$ includes $\tau $-reparametrizations as one of its local
transformations, and therefore it can be regarded as a generalization of
gravity on the worldline. The Sp$(2,R)$ gauge theory is non-trivial and
physically consistent only if the zero brane has two timelike coordinates $
X^0\left(\tau \right) ,X^{0^{\prime }}\left( \tau \right) $ in target space,
and has a global symmetry SO$(d,2)$, which is the Lorentz group with two
times.

Various gauge choices produce an infinite number of sectors of
one-time-physics, including free relativistic or non-relativistic particles
with or without mass, hydrogen atom, harmonic oscillator, particles in
various curved spacetimes (such as anti de Sitter space and others), and
even particles in arbitrary potentials $V\left( r\right) $. This is possible
because in the two-time theory there are an infinite number of ways to make
a gauge choice that defines physical ``time'' as known in one-time-physics.
All these sectors are connected to each other by Sp$(2)$ gauge
transformations (duality). The quantum Hilbert space for each of these
systems provide realizations of the SO$(d,2)$ symmetry, which is recognized
as the conformal symmetry in massless systems, and has other interpretations
in other systems, from the point of view of one-time-physics.

The SO$(d,2)$ group theoretical aspects are especially interesting in the
quantum theory. For diverse one-time physical sectors SO$(d,2)$ is realized
in the same unique representation, with the {\it same} eigenvalues of the
Casimir operators. Each one-time physical system provides a different basis
within the same representation of SO$(d,2)$, while duality transformations
are unitary transformations from one basis to another.

The two-time system has been generalized to include spin $\frac n2$ by
considering the gauge supergroup OSp$(n/2)$ \cite{dualsusy}. This
generalizes supergravity with $n$ supercharges on the worldline to a theory
with a larger local symmetry. Multi-particles are treated by using Sp$(2N)$
(instead of Sp$(2)$) and its supersymmetric generalizations \cite{sp2N} and
this makes a connection with the multiparticle formalism in \cite{sparticles}
.

Open problems include generalizing to target spacetime supersymmetric
version, investigating the two-time system in the presence of interactions
with background gravitational and gauge fields \cite{dualconf}, formulating
the second quantized version of the two-time theory, and generalizing the
scheme to strings or p-branes.

The main message is that two-time-physics is not only possible, but is a
basis for unifying many features of one-time-physics in a geometrical manner.

\section{Sp$\left( 2,R\right) $ gauge theory}

In the 1970's we gradually learned that what used to be considered global
symmetries became part of local gauge symmetries that unified all
interactions. There is a global symmetry that was not included in the
unification scheme, namely the Sp$\left( 2,R\right) $ global symmetry of all
Quantum Mechanics. Sp$\left( 2,R\right) $ treats generalized position and
momentum $(x,p)$\ as a doublet in phase space. The basic idea put forward by
BDA\cite{dualconf} is to turn this global symmetry into the gauge symmetry
of a theory\footnote{%
Historically, the Sp$\left( 2,R\right) $ gauge theory formalism gradually
developed from formalism that was used to construct ghost free two-time
models \cite{ibkounnas}{-}\cite{sparticles}. In turn these were motivated by
general supersymmetric structures that emerged in trying to understand
duality in M-theory and Super Yang-Mills theory, which provided hints for
two-times and higher dimensions from various points of view. However, the
idea can be stated as a principle which can be pursued quite independently
than the historical steps that led to it.}.

So far this idea has been used to construct the simplest model involving
particles (zero-branes), but the basic idea is more general and one may look
forward to applying it to more general cases including p-branes.

It is worth noting that a common factor in all duality transformations is a
transformation which mixes canonical conjugates and that belongs to Sp$
\left( 2,R\right) $. This is already the case in the oldest example of
Maxwell duality that acts between the canonical pairs of the electric and
magnetic fields $\left( {\bf E,B}\right) $ or electric-magnetic charges $
\left( e,g\right) $. Similarly there is an Sp$\left( 2,R\right) $ in the
Seiberg-Witten duality in a supersymmetric Yang-Mills theory. Furthermore,
T-duality in string theory transforms Kaluza-Klein momenta with winding
numbers in position space. Finally, the more general U-duality transforms
canonically conjugate electric-magnetic quantum numbers of p-branes. In all
these cases duality is a gauged discrete group which should be contrasted
with our continuously gauged Sp$\left( 2,R\right) $. Although these
dualities are discrete gauge transformations, it is not excluded that they
may arise from a more general continuous gauge theory after some gauge
fixing. In fact, examples of the discrete duality after gauge fixing exist
in our model with zero-branes: some of the one-time-physics models listed in
the introduction are related to each other by discrete Sp$\left( 2,R\right) $
transformations at fixed time in a Hamiltonian formalism.. These discrete
transformations are part of the continuous Sp$\left( 2,R\right) $.

A consequence of the Sp$\left( 2,R\right) $ gauge theory is that duality and
two-times are inextricably connected to each other. In fact, local Sp$\left(
2,R\right) $ symmetry requires one extra timelike coordinate plus one extra
spacelike coordinate to lift a system from one-time physics to its most
symmetric SO$\left( d,2\right) $ covariant form in two-time-physics. The
requirement of the extra dimensions to exhibit a higher symmetry is
consistent with similar observations involving duality in M-theory and its
extensions. In particular it is worth noting that (i) Type-IIA $
\longleftrightarrow $ 11D supergravity duality lifts 10D string theory to
M-theory in 11 dimensions, (ii) the 11D superalgebra (2-brane + 5-brane) of
M-theory is automatically a 12D superalgebra with signature $\left(
10,2\right) $, (iii) Type IIA$\longleftrightarrow $ Type IIB dualities lead
to F-theory in 12 dimensions, and (iv) requiring a sufficiently large
structure to unify TypeIIA + TypeIIB superalgebras leads to S-theory in 14
dimensions. By now we have become more accustomed to the idea that the
fundamental theory may take its most symmetric form when formulated in
higher dimensions. In fact the construction of the elusive fundamental
theory may first require a deeper understanding of the relationship between
one-time physics and the formulation of physics in higher dimensions with
more timelike coordinates.

In the remainder of this section we review the construction for zero-branes
given by BDA \cite{dualconf}. The theory is based on turning the global Sp$
(2,R)$ automorphism symmetry of the commutation relations in Quantum
Mechanics into a local symmetry of an action. Sp$\left( 2,R\right) $ treats
position and momentum $(x,p)$\ as a doublet in phase space. Consider the
particle (zero-brane) described by $X^{M}\left( \tau \right) $ and its
canonical conjugate $P^{M}\left( \tau \right) $. The signature of target
spacetime $\eta _{MN}$ and its relation to ordinary spacetime will be
determined below. To remove the distinction between position and momentum we
rename them $X_{1}^{M}\equiv X^{M}$ and $X_{2}^{M}\equiv P^{M}$ and define
the doublet $X_{i}^{M}=\left( X_{1}^{M},X_{2}^{M}\right) .$ The local Sp$
\left( 2,R\right) $ acts as follows
\begin{equation}
\delta _{\omega }X_{i}^{M}\left( \tau \right) =\varepsilon _{ik}\omega
^{kl}\left( \tau \right) X_{l}^{M}\left( \tau \right) .  \label{doublet}
\end{equation}
Here $\omega ^{ij}\left( \tau \right) =\omega ^{ji}\left( \tau \right) $ is
a symmetric matrix containing three local parameters of Sp$\left( 2,R\right)
$, and $\varepsilon _{ij}$ is the Levi-Civita symbol that is invariant under
Sp$\left( 2,R\right) $ and serves to raise or lower indices. The Sp$\left(
2,R\right) $ gauge field $A^{ij}\left( \tau \right) $ is symmetric in $(ij)$
and transforms in the standard way $\delta _{\omega }A^{ij}=\partial _{\tau
}\omega ^{ij}+\omega ^{ik}\varepsilon _{kl}A^{lj}+\omega ^{jk}\varepsilon
_{kl}A^{il}.$ The covariant derivative is $D_{\tau }X_{i}^{M}=\partial
_{\tau }X_{i}^{M}-\varepsilon _{ik}A^{kl}X_{l}^{M}.$

An action that is invariant under Sp$\left( 2,R\right) $ gauge symmetry is
\begin{eqnarray}
S_{0} &=&\frac{1}{2}\int_{0}^{T}d\tau \left( D_{\tau }X_{i}^{M}\right)
\varepsilon ^{ij}X_{j}^{N}\eta _{MN}  \label{action} \\
&=&\int_{0}^{T}d\tau \left( \partial _{\tau }X_{1}^{M}X_{2}^{N}-\frac{1}{2}
A^{ij}X_{i}^{M}X_{j}^{N}\right) \eta _{MN}\,\,.  \nonumber
\end{eqnarray}
where we have dropped a total derivative $\partial _{\tau }\left( -\frac{1}{%
2 }X_{1}\cdot X_{2}\right) $ from the Lagrangian. The canonical conjugates
are $X_{1}^{M}=X^{M}$ and $\partial S/\partial \dot{X}%
_{1}^{M}=X_{2}^{M}=P^{M}$. They are consistent with the idea that $%
(X_{1}^{M},X_{2}^{M})$ is the doublet $\left( X^{M},P^{M}\right) $. The
equations of motion for $A^{ij}$ that follows from the Lagrangian (\ref%
{action}) give the first class constraints
\begin{equation}
X\cdot X=X\cdot P=P\cdot P=0.  \label{constraints}
\end{equation}
Their Lie algebra is Sp$\left( 2,R\right) $. If the signature of $\eta _{MN}$
corresponds to a single time-like coordinate the only classical solution of
the constraints is that $X^{M},P^{M}$ are parallel and lightlike. This is
trivial in the sense that there is no angular momentum. Non-trivial
solutions are possible provided the signature of $\eta _{MN}$ corresponds to
two time-like coordinates. More timelike coordinates are not allowed because
the gauge symmetry is insufficient beyond two timelike dimensions to remove
ghosts.

The action has a global symmetry under global Lorentz transformations SO$
\left( d,2\right) $ which leave the metric $\eta _{MN}$ invariant. The
generators of this symmetry are
\begin{equation}
L^{MN}=\varepsilon ^{ij}X_i^MX_j^N=X^MP^N-X^NP^M.  \label{sod2}
\end{equation}
These $L^{MN}$ are gauge invariant under Sp$\left( 2,R\right) $ for each $
M,N $. Other gauge invariants include $\varepsilon ^{ij}X_i^MD_\tau
X_j^N,\,\varepsilon ^{ij}D_\tau X_i^MD_\tau X_j^N$, etc., but these vanish
on shell as a result of the equations of motion $D_\tau X_i^M=0$.

The full {\it physical information of the theory is contained in the gauge
invariant }$L^{MN}$. Using the constraints (\ref{constraints}) it is
straightforward to show that all the Casimir operators of SO$\left(
d,2\right) $ vanish at the classical level
\begin{equation}
Classical:\quad C_{n}\left( SO\left( d,2\right) \right) =\frac{1}{n!}
Tr\left( iL\right) ^{n}=0.
\end{equation}
In the first quantized theory the quantum states are labelled by both Sp$
\left( 2,R\right) $ and SO$\left( d,2\right) $ Casimir eigenvalues in the
form $|C_{2}\left( Sp(2,R)\right) ;C_{n}\left( SO\left( d,2\right) \right) >$
since the generators of these groups commute, and we need to find their
eigenvalues for physical states. Generally the possible $Sp(2,R)$ quantum
numbers are $|jm>$. In contrast to the classical theory, the quantized $
C_{n}\left( SO\left( d,2\right) \right) $ are not zero after taking quantum
ordering into account. The following relations are proven by writing out all
the Casimir operators in terms of $X,P$ while respecting their order. First,
all Casimir eigenvalues $C_{n}\left( SO\left( d,2\right) \right) $ are
rewritten in terms of $C_{2}\left( SO\left( d,2\right) \right) $ and $d$.
For example $C_{3}\left( SO\left( d,2\right) \right) =\frac{d}{3!}
C_{2}\left( SO\left( d,2\right) \right) ,$ etc.. Second, the quadratic
Casimir of Sp$\left( 2,R\right) $ is related to the quadratic Casimir of SO$
\left( d,2\right) $ by $C_{2}\left( SO\left( d,2\right) \right)
=4C_{2}\left( Sp(2,R)\right) +1-\frac{d^{2}}{4}$. Third, since physical
states are gauge invariant, the quadratic Casimir of Sp $\left( 2,R\right) $
must vanish in the physical sector (i.e. $j=0$ and $m=0$). The last
condition fixes all the Casimir eigenvalues for SO$\left( d,2\right) $ to
unique non-zero values in terms of $d$. Therefore the quantum system can
exist only in a unique unitary representation of $SO\left( d,2\right) $
characterized by \cite{dualconf}
\begin{equation}
Quantum:\quad C_{2}(Sp\left( 2\right) )=0,\quad \left\{
\begin{array}{l}
C_{2}\left( SO\left( d,2\right) \right) =1-\frac{d^{2}}{4}, \\
C_{3}\left( SO\left( d,2\right) \right) =\frac{d}{3!}\left( 1-\frac{d^{2}}{4}
\right) \\
\cdots%
\end{array}
\right.  \label{covquant}
\end{equation}
This information completely specifies the physical sector of the Hilbert
space in a unique $SO\left( d,2\right) $ representation. It should be
possible to obtain a similar result by using the methods of BRST
quantization \cite{jarvis} but this remains to be done.

From this gauge invariant result it follows that the diverse
one-time-physics models that emerge by gauge fixing must have precisely zero
SO$\left( d,2\right) $ Casimir eigenvalues at the classical level, and also
the same non-trivial Casimir eigenvalues (\ref{covquant}) that label the
unique physical Hilbert space in their first quantized versions. This is, of
course, a natural result of the formalism, however the prediction it makes
for diverse one-time-physics systems was not recognized to be true before,
and seems to be amazing. For example it suggests that the free relativistic
massless particle in $\left( d-1\right) $ space dimensions should have a
Hilbert space dual to that of the particle moving in the $1/r$ potential in
the same number of dimensions, and (except for the choice of basis) should
be described by the same unique SO$\left( d,2\right) $ representation, etc..
This prediction of two-time-physics has been explicitly verified\cite{dualH}
{-}\cite{dualsusy} to be correct not only for this example, but for many
other cases as well, including spinning particles (for which the Casimir
eigenvalues change according to the value of the spin). The fact that this
test succeeded is encouraging evidence for two-time-physics.

\section{Gauge choices, dual physics}

First we describe in general terms how diverse one-time physical systems
emerge from the same two-time theory by taking various gauge choices that
embed physical time in different ways in the extra dimensions. It is evident
that, by the very procedure in which they are derived, these diverse
physical systems are Sp$\left( 2,R\right) $ duals of each other.

We have the freedom to fix up to 3 functions since Sp$\left( 2\right) $ has
3 gauge parameters. The procedure is as follows. (1) Make $n$ gauge choices
(using $n$=2 or 3) for some of the $2d+4$ functions $X^M\left( \tau \right)
,P^M\left( \tau \right) $. (2) Solve $n$ constraints which determines $n\,$
additional functions, thus obtaining a gauge fixed configuration $
X_0^M\left( \tau \right) ,P_0^M\left( \tau \right) $ parametrized in terms
of $2\left( d+2-n\right) $ independent functions $x\left( \tau \right)
,p\left( \tau \right) $.(3) The dynamics for the remaining degrees of
freedom $x,p$ is determined by inserting $X_0^M\left( \tau \right)
,P_0^M\left( \tau \right) $ in the original action (\ref{action}), thus
constructing a new one-time-physics action \footnote{%
Thus, for $d=4$, if we choose all the gauges in a particular way and solve
all the constraints the remaining three positions and three momenta
correspond to ordinary 3D physical phase space ${\bf r,p}$.}
\begin{equation}
S\left( x,p\right) =S_0\left( X_0^M,P_0^M\right) =\int d\tau \,\,L\left(
x\left( \tau \right) ,p\left( \tau \right) ,A\left( \tau \right) \right) .
\end{equation}
Here $A\left( \tau \right) $ is a remaining gauge potential if $n=2,$ but $
A\left( \tau \right) $ is absent if $n=3$ (the $A$'s in eq.(\ref{action})
drop out when the corresponding constraint is solved explicitly). The
one-time physical system that emerges is recognized by studying the form of
the Lagrangian $L$.

In this approach it is no surprise to find that $S\left( x,p\right) $
inherits the SO$\left( d,2\right) $ symmetry, which however is now realized
non-linearly. The presence of this hidden symmetry was {\it not suspected
for most of the diverse physical systems} constructed by this procedure,
although it was known before for a couple of examples (free relativistic
particle, and hydrogen atom). Since the original generators of the symmetry $
L^{MN}$ in (\ref{sod2}) are gauge invariant, they must be the generators of
the symmetry of the new action. Indeed the symmetry generators for the new
action can be constructed easily at any $\tau $ by inserting $X_0^M\left(
\tau \right) ,P_0^M\left( \tau \right) $ in the $L^{MN}$ of eq.(\ref{sod2})
\begin{equation}
L^{MN}\left( x\left( \tau \right) ,p\left( \tau \right) ,\tau \right)
=X_0^M\left( \tau \right) P_0^N\left( \tau \right) -X_0^N\left( \tau \right)
P_0^M\left( \tau \right) .  \label{lmnclass}
\end{equation}
It can be checked explicitly that these $L^{MN}$ form the algebra of SO$
\left( d,2\right) $ under Poisson brackets by using the fundamental Poisson
brackets for $\left\{ x,p\right\} =\eta $, (where $\eta =\pm 1$ depending on
the spacelike/timelike nature of the canonical variables $x,p$)
\begin{equation}
\left\{ L^{MN},L^{RS}\right\} =\eta ^{MR}L^{NS}+\eta ^{NS}L^{MR}-\eta
^{NR}L^{MS}-\eta ^{MS}L^{NR}.  \label{sod2alg}
\end{equation}
Explicit $\tau $ dependence\footnote{%
In the $n=2$ case there is a canonical time coordinate $t\left( \tau \right)
$ that is canonically conjugate to a momentum $p_t\left( \tau \right) $. The
explicit $\tau $ appears when the $n=2$ case is gauge fixed to the $n=3$
case by taking $t\left( \tau \right) =\tau $ while $p_t$ is solved from the
remaining constraint as the Hamiltonian that depends on the remaining
canonical variables. The Hamiltonian is independent of $\tau $, but some of
the generators $L^{MN}$ are functions of $\tau $ in addition to $x\left(
\tau \right) ,p\left( \tau \right) $. The explicit $\tau $ is part of a
local gauge transformation that corresponds to $\tau $ reparametrization
needed to maintain the form of the gauge after a naive SO$\left( d,2\right) $
transformation is applied on the coordinates. The same situation arises in
the familiar case of the relativistic particle with action $\int d\tau \sqrt{
-\dot{x}^2}$.} appears in the new $L^{MN}$ when $n=3\ $(but not when $n=2$).
If $\tau $ appears explicitly it is treated as a parameter in the Poisson
brackets as opposed to a canonical degree of freedom $x\left( \tau \right)
,p\left( \tau \right) $. Then the SO$\left( d,2\right) $ algebra holds for
all $\tau $. The symmetry transformations of the canonical coordinates $
\delta x\left( \tau \right) ,\delta p\left( \tau \right) $ are obtained by
evaluating the Poisson brackets
\begin{equation}
\delta x\left( \tau \right) {\bf =}\frac 12\varepsilon _{MN}\left\{
L^{MN}\left( \tau \right) ,x\left( \tau \right) \right\} ,\quad \delta
p\left( \tau \right) {\bf =}\frac 12\varepsilon _{MN}\left\{ L^{MN}\left(
\tau \right) ,p\left( \tau \right) \right\} .,
\end{equation}
while continuing to treat $\tau $ as a parameter. The explicit $\tau $
dependence in these transformation laws is vital for demonstrating the
symmetry of the action when $n=3$. Indeed one can verify explicitly that the
one-time-physics action $S\left( x,p\right) $ is invariant under SO$\left(
d,2\right) $ because the Lagrangian transforms like a total derivative at
any $\tau $
\begin{equation}
\delta L=\partial _\tau \Lambda \left( \tau ,\varepsilon _{MN}\right) .
\end{equation}
It must be emphasized that we obtain invariance of the {\it action} under SO$
(d,2)$, which must be distinguished from invariance of the {\it Hamiltonian}
. This distinction arises when all three gauge choices are made and a
Hamiltonian is defined. Recall that in that case the generators of the
symmetry depend explicitly on $\tau$ as emphasized before, which means they
are conserved in the sense that the total time derivative, including
derivative with respect to the explicit $\tau$, is zero. On the other hand,
the generators that commute with the Hamiltonian are only those that do not
depend on $\tau$ explicitly \footnote{%
This is a familiar phenomenon. For example for the standard relativistic
particle the action is invariant under rotations as well as boosts, but the
Hamiltonian defined after gauge fixing $x^0(\tau)=\tau$ is invariant only
under rotations.}. In this sense SO$(d,2)$ should be understood as the
dynamical symmetry of the system. For example, for the H-atom, the symmetry
of the Hamiltonian is SO$(d)$ while the dynamical symmetry is SO$(d,2)$. We
will see below more examples of physical systems all of which have SO$(d,2)$
dynamical symmetry (equivalently symmetry of the action) as a trivial
consequence of our formulation, but which was not known and was unexpected
before for those systems.

The one-time physical system described by $L(x,p)$ can be first quantized in
the usual way. One may then construct the quantum generators of SO$\left(
d,2\right) $ from the classical ones (\ref{lmnclass}). In a Hamiltonian
formalism we take $\tau =0$ , and order the operators to insure hermitian $
L^{MN}\left( x,p\right) $. In this procedure we find that we also need
corrections of some $L^{MN}$ by including some anomaly terms to insure
closure of the SO$\left( d,2\right) $ algebra at the quantum level (i.e.
orders of operators respected). Once this is achieved in some fixed gauge we
know that the physical space for the corresponding physical system is
described by a representation of the SO$\left( d,2\right) $ algebra. The
remaining question is whether the representation is the same one as the one
specified in covariant quantization in eq.(\ref{covquant}). Indeed we find
complete agreement in every case in which this procedure has been carried
out. This includes the free relativistic massless particle, the hydrogen
atom, the harmonic oscillator, the particle moving on AdS$_{d-n}\times S^n$
and a few of these cases including spinning particles (for which the
original action and the Casimirs are generalized to include the effects of
spin).

\section{Examples}

We will give here some examples to illustrate the ideas and procedures
described above. More cases including spinning generalizations are available
in the literature.

\subsection{Free massless particle}

We use the basis $X^M=\left( X^{+^{\prime }},X^{-^{\prime }},x^{\mu
\,\,}\right) $ with the metric $\eta ^{MN}$ taking the values $\eta
^{+^{\prime }-^{\prime }}=-1$ and $\eta ^{\mu \nu }=$Minkowski. We choose 2
gauges $X^{+^{\prime }}=1,\,P^{+^{\prime }}=0$, and solve 2 constraints $
X^2=X\cdot P=0$
\begin{eqnarray*}
M &=&\left( +^{\prime },\quad -^{\prime }\quad ,\,\mu \right) \\
X_0^M\left( \tau \right) &=&\left( 1,\frac{x^2\left( \tau \right) }2,x^{\mu
\,}\left( \tau \right) \right) , \\
P_0^M\left( \tau \right) &=&\left( 0,\,p\left( \tau \right) \cdot x\left(
\tau \right) \,,p^{\mu \,\,}\left( \tau \right) \right) .
\end{eqnarray*}
Inserting these in the action \ref{action}, we find
\begin{eqnarray}
S\left( x,p\right) &=&\int_0^Td\tau \left( -\dot{X}^{+^{\prime
}}P^{-^{\prime }}-\dot{X}^{-^{\prime }}P^{+^{\prime }}+\dot{x}^{\mu
\,}p_{\mu \,\,}-\frac{A^{22}}2p^2-0-0\right)  \nonumber \\
&=&\int_0^Td\tau \left( \dot{x}^{\mu \,}p_{\mu \,\,}-\frac
12A^{22}p^2\right) \Rightarrow \frac 12\int_0^Td\tau \frac{\dot{x}^2}{A^{22}}
\label{masslessrel}
\end{eqnarray}
which is obviously interpreted as the action for the massless relativistic
particle.

A third gauge choice can be made by taking $x^{+}\left( \tau \right) =\tau $
and then solving the constraint $P^2=0$ which gives $p^{-}=\vec{p}^2/2p^{+}$
namely the Hamiltonian in the lightcone gauge. Inserting these either in the
original action (\ref{action}) or in the intermediate action (\ref%
{masslessrel}) produces the action for the remaining independent degrees of
freedom $\left( x^{-},p^{+}\right) $ and $\left( \vec{x},\vec{p}\right) $
\begin{equation}
S\left( x,p\right) =\int_0^Td\tau \left( -\partial _\tau x^{-}p^{+}-\partial
_\tau \vec{x}\cdot \vec{p}-\vec{p}^2/2p^{+}\right) .  \label{masslessLC}
\end{equation}

The gauge invariant observables are the $L^{MN}$. They can be constructed
explicitly as described above\cite{dualconf}{-}\cite{dualH} either for the
action (\ref{masslessrel}) or the action (\ref{masslessLC}). One can easily
compute $\delta x\left( \tau \right) $ and $\delta p\left( \tau \right) $
and verify explicitly that all forms of the action are symmetric under all SO%
$\left( d,2\right) $ transformations. Note that the transformations
generated by $L^{MN}$ for the action (\ref{masslessrel}) are independent of $%
\tau$ ($n=2$ case) but those for the action (\ref{masslessLC} ) depend on $%
\tau $ ($n=3$ case).

In a Hamiltonian formalism at $\tau =0$ the first quantized generators for
either action (\ref{masslessrel}) or (\ref{masslessLC}) have anomalous terms
due to quantum ordering, which is necessary for hermiticity and closure of
the algebra. Using the corrected quantum generators it has been verified\cite%
{dualconf}{-}\cite{dualH} that the quadratic Casimir is precisely $
C_2=\left( 1-d^2/4\right) $, in agreement with the prediction in eq.(\ref%
{covquant}). This result is also obtained in the field theory version of the
free massless particle\cite{dualconf}.

\subsection{Particle in a potential $V\left( r\right) $}

In this section we follow the ideas of \cite{dualH}{-}\cite{dualsusy}. We
use the basis $X^{M}=\left( X^{0^{\prime }},X^{0},X^{I}\right) $, with
metric $\eta ^{0^{\prime }0^{\prime }}=\eta ^{00}=-1$ and $\eta ^{IJ}=\delta
^{IJ}$. The Sp$\left( 2\right) $ gauge symmetry permits us to make three
gauge choices and then solve the three constraints $X^{2}=P^{2}=X\cdot P=0.$
This eliminates six functions from phase space or expresses them in terms of
independent degrees of freedom. In \cite{dualH}{-}\cite{dualsusy} the
following choices of gauge and solution of constraints was given
\begin{eqnarray}
M &=&\left[ \,\,\,0^{\prime }\,\,\,\,\,\,\,\,,\quad 0\quad \,\,\,\,,\quad
\quad \quad 1^{\prime }\quad \quad ,\quad \quad \quad \quad i\quad \quad
\quad \quad \quad \right]   \nonumber \\
X^{M} &=&\left[ \cos u,\,\,-\sin u\,\,\,,-\,{\bf \hat{R}\cdot P}\frac{\sqrt{%
-2H}}{V}{\bf ,\quad }\,({\bf \hat{R}}^{i}+\frac{1}{V}{\bf \hat{R}\cdot PP}%
^{i}{\bf )}\right] \frac{VR}{\gamma },  \label{hatomgauge} \\
P^{M} &=&\left[ \,\sin u\,\,,\,\,\cos u\,\,\,,\quad (1+\frac{{\bf P}^{2}}{V}%
)\quad ,\quad \quad \sqrt{-2H}\,\frac{1}{V}{\bf P}^{i}{\bf \,\,\,\,}\right]
\frac{\gamma }{\sqrt{-2H}}.
\end{eqnarray}%
The independent degrees of freedom are ${\bf R,P}$. Here $V\left( {\bf R,P}%
\right) $ is any function of the canonical variables while $H$ is the
Hamiltonian
\begin{equation}
H=\frac{{\bf P}^{2}}{2}+V<0.
\end{equation}%
To determine which class of potentials $V\left( R\right) $ is possible we
consider the generators of SO$\left( d,2\right) ~L^{MN}$ which must be
conserved in any gauge since they correspond to the global symmetries of the
action. Since both the action and the $L^{MN}$ are gauge invariant, the $%
L^{MN}$ expressed in any gauge must be conserved when we use the equations
of motion that follow from the gauge fixed action. Hence consider $%
L^{0^{\prime }0}$ which becomes in this gauge%
\begin{equation}
L^{0^{\prime }0}=X^{0^{\prime }}P^{0}-X^{0}P^{0^{\prime }}=\frac{RV}{\sqrt{%
-2H}}
\end{equation}%
Since $H$ is guaranteed to be a constant of motion, the remaining
possibility is that $RV$ must be a constant number, or more generally a
constant of motion. Let us first consider the case $V=-\alpha /R$ where $%
\alpha $ is a positive constant. In that case we further make the additional
gauge choice
\begin{equation}
u\left( \tau \right) =\left( {\bf r\cdot p}-2\tau H\right) \frac{\sqrt{-2H}}{%
\alpha }.
\end{equation}%
We also take $\gamma $ a constant, although $\gamma $ plays no role in the
classical theory because it drops out in all gauge invariant expressions,
but it can play a role in the quantum theory due to quantum ordering as
discussed in \cite{dualsusy}. In this form we have expressed the original $%
\left( d+2\right) $ degrees of freedom $\left( X^{M},P^{M}\right) $ in terms
of $\left( d-1\right) $ independent canonical degrees of freedom $\left(
{\bf R}^{i},{\bf P}^{i}\right) $.

Inserting these expressions in the original gauge invariant action gives the
dynamics for the independent degrees of freedom $\left( {\bf R,P}\right) $.
Since the constraints are already solved we get (see \cite{dualH}-\cite%
{dualsusy})
\begin{eqnarray}
S &=&\int_0^Td\tau \,\,\left( \partial _{\tau \,}X_1^MX_2^N\,\eta
_{MN}-0-0-0\right) \\
&=&\int_0^Td\tau \,\,\int_0^Td\tau \,\,\left( {\bf P}^i\partial _{\tau \,}
{\bf R}^i\,{\bf -}H\right) ,  \nonumber
\end{eqnarray}
where $H$ is identified as the Hamiltonian given above. In addition, the
last line shows that the $({\bf R,P}$) used above are indeed canonical
conjugates.

\subsubsection{Positive energies}

A similar gauge exists for a positive Hamiltonian. It is obtained by
exchanging the roles of $X^{0}$ and $X^{1^{\prime }}$ as follows, which is
equivalent to an analytic continuation from $H<0$ to $H>0$ in all
expressions. Thus, consider the gauge choice and solutions of constraints
given by
\begin{eqnarray}
M &=&\left[ \,\quad 0^{\prime }\quad \,\,\,\,,\quad \quad 1^{\prime }\quad
,\,\,\quad \quad \,0\,\,\,\,\,\,\,\,\quad ,\quad \quad \quad \quad \quad
i\quad \quad \quad \right]   \nonumber \\
\tilde{X}^{M} &=&\left[ \,\,\cosh w\,\,\,,\,\,\sinh w\,{\bf ,}%
\,\,\,\,\,\,\,\,\,\,\,{\bf \hat{r}\cdot \hat{p}}\frac{\sqrt{2H}}{V},\,{\bf %
\quad }\,({\bf \hat{r}}^{i}+\frac{1}{V}{\bf \hat{r}\cdot pp}^{i}{\bf )}%
\right] \frac{VR}{\gamma }, \\
\tilde{P}^{M} &=&\left[ \,\sinh w\,\,\,,\quad \cosh w\quad ,\,\quad (1+\frac{%
{\bf p}^{2}}{V})\quad \,\,,\,\,\quad \frac{\sqrt{2H}}{V}\,{\bf p}^{i}{\bf %
\,\,\,\,}\right] \frac{\gamma }{\sqrt{2H}},
\end{eqnarray}%
and
\begin{equation}
w=\left( {\bf r\cdot p}+2\tau H\right) \frac{\sqrt{2H}}{\alpha },
\end{equation}%
with
\begin{equation}
H=\frac{{\bf p}^{2}}{2}+V\left( {\bf r,p}\right) >0.
\end{equation}%
These expressions are related to the previous ones by the analytic
continuation $\sqrt{-2H}\rightarrow i\sqrt{2H}$, $u\rightarrow iw$, and then
switching $X^{0}$ and $X^{1^{\prime }}$ to eliminate the complex number $i$
(the factor of $i$ converts a spacelike coordinate to a timelike coordinate
and vice-versa). Inserting these gauge choices in the action we get again
\begin{eqnarray}
S &=&\int_{0}^{T}d\tau \,\,\left( \partial _{\tau
\,}X_{1}^{M}X_{2}^{N}\,\eta _{MN}-0-0-0\right)  \\
&=&\int_{0}^{T}d\tau \,\,\left( {\bf p}^{i}\partial _{\tau \,}{\bf r}^{i}%
{\bf -}H\right)   \nonumber
\end{eqnarray}%
Thus, in switching from negative to positive energies we need to make an
analytic continuation which is equivalent to an Sp$\left( 2\right) $ gauge
transformation $\left( X^{M},P^{M}\right) \rightarrow \left( \tilde{X}^{M},%
\tilde{P}^{M}\right) $. Hence the canonical conjugates $\left( {\bf r,p}%
\right) $ used for positive energies must be related to $\left( {\bf R,P}%
\right) $ used for negative energies by a local Sp$\left( 2\right) $ gauge
transformation.

\subsubsection{SO$\left( d\right) $ or SO$\left( d-1,1\right) $ symmetric
Hamiltonians}

Next we discuss the symmetries of $H$ (as opposed to the symmetries of the
action $S$) for a special class of Hamiltonians. For an arbitrary potential $
V$ the evident symmetry is rotation symmetry SO$\left( d-1\right) $. Next
consider a potential $V$ of the form $V=-\frac \alpha R.$ If $\alpha $ is a
constant this Hamiltonian describes the H-atom or the Kepler problem. For
this case it is well known that $H$ has a hidden SO$\left( d\right) $
symmetry. However, one may go beyond a constant $\alpha $ and still have SO$
\left( d\right) $ symmetry$.$ In particular if we choose $\alpha $ to be a
function of $H$ then the SO$\left( d\right) $ symmetry is preserved. That
is, if the Hamiltonian \thinspace $H$ is solved from the equation
\begin{equation}
H=\frac{{\bf P}^2}2-\frac{\alpha \left( H\right) }R
\end{equation}
for any function $\alpha \left( H\right) $, the resulting $H\left( {\bf R,P}
\right) $ will be shown to have SO$\left( d\right) $ symmetry. This is seen
by constructing the $L^{MN}$ as described above. The SO$\left( d,2\right) $
generators are (classical, and $\tau =0$)
\begin{eqnarray*}
L^{0^{\prime }I} &=&\frac{\alpha \left( H\right) }{\sqrt{-2H}}m^I,\quad
L^{0I}=-\frac{\alpha \left( H\right) }{\sqrt{-2H}}n^I \\
L^{0^{\prime }0} &=&\frac{\alpha \left( H\right) }{\sqrt{-2H}},\quad L^{IJ}=
\frac{\alpha \left( H\right) }{\sqrt{-2H}}\left( n^Im^J-n^Jm^I\right) .
\end{eqnarray*}
where $n^I,m^I$ are unit vectors that are orthogonal (as seen from (\ref%
{hatomgauge})). The SO$\left( d\right) $ generators $L^{IJ}$ include
rotations $L^{ij}$ and the $L^{1^{\prime }i}$=Runge-Lenz vector $
L^{1^{\prime }i}$ in $d-1$ dimensions
\[
L^{ij}={\bf R}^i{\bf P}^j-{\bf R}^j{\bf P}^i,\,\,\,\quad L^{1^{\prime }i}=
\frac{\alpha \left( H\right) }{\sqrt{-2H}}\left( \frac 12L^{ij}{\bf P}
_j+\frac 12{\bf P}_jL^{ij}-\frac{\alpha \left( H\right) {\bf R}^i}R\right) .
\]
We see that both the SO$\left( d\right) $ quadratic Casimir and the SO$
\left( d\right) $ singlet generator $L^{0^{\prime }0}$ are functions of only
$H$ at the classical level
\begin{eqnarray*}
C_2\left( SO\left( d\right) \right) &=&\left( L^{1^{\prime }i}\right)
^2+\frac 12\left( L^{ij}\right) ^2=\frac{\alpha ^2\left( H\right) }{-2H}, \\
\left( L^{0^{\prime }0}\right) ^2 &=&\frac{\alpha ^2\left( H\right) }{-2H}.
\end{eqnarray*}
Of course the relation between $H$ and the Casimir $C_2\left( SO\left(
d\right) \right) $ receives quantum corrections but the relation between $
L^{0^{\prime }0}$ and $H$ remains the same despite quantization. Furthermore
$H$ is obviously invariant under $SO\left( d\right) $ since $L^{0^{\prime
}0} $ commutes with all the SO$\left( d\right) $ generators $L^{IJ}$.

To do the quantum mechanics it is sufficient to use that $H$ is a function
of the generator $L^{0^{\prime }0}$. Since the three generators $
L^{0^{\prime }0},L^{0^{\prime }1^{\prime }},L^{1^{\prime }0}$ form an SO$
\left( 1,2\right) $ algebra we can immediately determine group theoretically
the spectrum of $H$ from the spectrum of $L^{0^{\prime }0}$. This procedure
was followed in \cite{dualH} to discuss the spectrum of the H-atom, and now
it can be generalized to the more general case $\alpha \left( H\right) $ in
a straightforward manner. Using the permitted eigenvalues of $L^{0^{\prime
}0}$ as determined in \cite{dualH}, $L^{0^{\prime }0}=(\frac 12\left(
d-4\right) +n)$, with $n=0,1,2,\cdots $, the spectrum is obtained from the
relation between $L^{0^{\prime }0}$ and $H$
\begin{equation}
E_n=-\frac{\alpha ^2\left( E_n\right) }{2(\frac 12\left( d-4\right) +n )^2}.
\end{equation}
Thus, this method provides a new class of exactly solvable Hamiltonians $
H\left( R,P\right) $ that have an SO$\left(d\right) $ symmetry, just like
the H-atom does. Since the excitation spectrum is now a different function
of $n$ (determined by the choice of $\alpha \left( H\right) $) such a
Hamiltonian may have interesting applications.

In summary, If $RV\left( {\bf R,P}\right) =\alpha \left( H\right) $ is a
function of $H\left( {\bf R,P}\right) $ we have argued that the Hamiltonian
is symmetric under SO$\left( d\right) $ symmetry for $H<0,$ and similarly
symmetric under an SO$\left( d-1,1\right) $ symmetry when $H>0.$ There may
be a bigger class of SO$\left( d\right) $ symmetric Hamiltonians that remain
to be found by using a more general gauge. The action $S\left( {\bf R,P}%
\right) $ is invariant under a dynamical symmetry SO$\left( d,2\right) $ as
described in the general discussion. When the Hamiltonian can be written as
a simple expression of the generators of SO$\left( d,2\right) $ the
dynamical symmetry can be used to provide a group theoretical solution of
the eigenstates, eigenenergies and other physical properties of the system.
A class of such simple physical systems is obtained when $\alpha \left(
H\right) $ is an arbitrary function of $H.$

\subsection{Particle in curved space - AdS$_{d-n}\times S^n$ gauge}

The example of a particle moving in an AdS$_{d-n}\times S^n$ background
should be of special current interest in view of the proposed $AdS-CFT$
duality \cite{maldacena}. We begin with $AdS_d$. Using the basis $X^M=\left(
X^{0^{\prime }},X^{1^{\prime }},X^0,X^i\right) $ we choose 2 gauges $
X^{1^{\prime }}=1,\,P^{1^{\prime }}=0$, and solve the 2 constraints $
X^2=X\cdot P=0$. The solution is parametrized as follows
\begin{eqnarray*}
M &=&\left( 0^{\prime }\quad \quad ,\quad \quad \quad 1^{\prime }\quad
,\quad \quad \quad 0\quad \quad ,\quad \quad i\quad \right) \\
X_0^M &=&\left( A(r)\cos t,\,\,\quad 1\,\,,\,\,\,\quad A(r)\sin
t,\,\,\,\quad B(r){\bf \hat{r}\,\,}\right) ,\,\,A^2-B^2=1, \\
P_0^M &=&\left(
\begin{array}{c}
-\frac{p^0}A\sin t \\
+\frac{AB}{\partial _rB}p_r\cos t%
\end{array}
,0,
\begin{array}{c}
\frac{p^0}A\cos t \\
+\frac{AB}{\partial _rB}p_r\sin t%
\end{array}
,\,
\begin{array}{c}
\frac rB({\bf p-}p_r{\bf \hat{r})} \\
+\frac{A^2}{\partial _rB}p_r{\bf \hat{r}}%
\end{array}
\right)
\end{eqnarray*}
Inserting this gauge in the action we find
\begin{eqnarray*}
S_0 &=&\int_0^Td\tau \,\,\left( \partial _{\tau \,}X_1^MX_2^N\,\eta
_{MN}-\frac 12A^{22}X_2\cdot X_2-0-0\right) \\
&=&\int_0^Td\tau \,\left( \dot{x}^{\mu \,}\cdot p_{\mu \,}-\frac
12A^{22}G^{\mu \nu }\left( x\right) \,p_{\mu \,}p_{\nu \,}\right)
\rightarrow \int_0^Td\tau \frac{G_{\mu \nu }\dot{x}^\mu \dot{x}^\nu }{
2A^{22} }
\end{eqnarray*}
where the last expression is obtained by using the equation of motion for $p$
or by integrating out $p$ in the path integral. This describes a particle
moving in a curved background. The metric $G_{\mu \nu }$ (lower indices) is
given by
\[
ds^2=dX\cdot dX=G_{\mu \nu }dx^\mu dx^\nu =-A^2dt^2+\frac{(\partial _rB)^2}{
A^2}dr^2+B^2d\Omega ^2.
\]
For example, $A=\sqrt{1+r^2},$ and $B=r$ gives
\[
\left( ds^2\right) _{AdS_d}=-\left( 1+r^2\right) \,dt^2+\frac
1{1+r^2}dr^2+r^2d\Omega ^2,
\]
which is recognized as a particular parametrization of anti de Sitter space $
AdS_d$ in $\left( d-1,1\right) $ dimensions. Another example with $A=\frac{
1+r^2}{1-r^2},$ and $B=\frac{2r}{1-r^2}$ gives
\[
\left( ds^2\right) _{AdS_d}=-\left( \frac{1+r^2}{1-r^2}\right) ^2dt^2+\left(
\frac{2r}{1-r^2}\right) ^2d{\bf r}^2.
\]
which is a different form of the $AdS_d$ metric.

To construct the particle in $AdS_{d-n}\times S^n$ we divide the $d+2$
components of $X^M$ into two sets, $X^M=(x_{d-n+1}^m,y_{n+1}^i)$. The first
set $x_{d-n+1}^m$ contains $d-n+1$ components that include the two timelike
dimensions and the second set $y_{n+1}^i$ contains $n+1$ components that are
purely spacelike. Similarly with $P^M$. Then we make the two gauge choices
\begin{equation}
y\cdot y=1,\quad y^ip_i=0.
\end{equation}
That is, $y$ is a unit vector while the corresponding radial component of $
p^i$ vanishes. This is to be compared to the $n=0$ case treated above.
Solving the constraints $X^2=X\cdot P=0$, and inserting the result in the
action (\ref{action}) we derive the action for the particle moving in a
curved background with the metric computed from $ds^2=dX\cdot dX=dx\cdot
dx+dy\cdot dy$. We find the $AdS_{d-n}\times S_n$ metric.
\begin{equation}
ds^2=ds_{AdS_{d-n}}^2+\left( d\Omega _n\right) ^2
\end{equation}
where $\Omega _n$ describes the $n$-sphere defined by the unit vector $
y_{n+1}^i$.

Just like all previous cases the action with this metric has an SO$\left(
d,2\right) $ symmetry. This contains hidden symmetries not noticed before
since SO$\left( d,2\right) $ is larger than the popularly known symmetry in
the $AdS_{d-n}\times S_n$ background. Namely the $AdS_{d-n}$ piece has an SO$
\left( d-n-1,2\right) $ symmetry and the $S^n$ piece has an SO$\left(
n+1\right) $ symmetry, while our approach shows that the overall system has
a larger symmetry SO$(d,2)$
\begin{equation}
SO\left( d,2\right) \supset SO\left( d-n-1,2\right) \times SO\left(
n+1\right) .
\end{equation}
For example, the action for a particle moving in $AdS_3$ has an SO$(3,2)$
symmetry, which is larger than the popularly known SO$(2,2)$. The action for
the particle moving on $AdS_5 \times S^5$ has an SO$(10,2)$ symmetry which
is larger than the expected symmetry SO$(4,2) \times$ SO$(6)$. Again, the
presence of the larger symmetry is the evidence for the presence of
two-time-physics. The symmetry transformations and the quantum generators
for this case are discussed elsewhere in more detail \cite{gauges}.

\subsection{Conformal factors}

Consider a gauge that has a solution $X_0^M\left( x,p\right) ,P_0^M\left(
x,p\right) $ which satisfies the constraints $X_0^2=X_0\cdot P_0=0,$ and
which gives $\dot{X}_0^M\cdot P_{0M}=\dot{x}^{\mu \,}\cdot p_{\mu \,}$ and $
P_0^2=G_0^{\mu \nu }\left( x\right) \,p_{\mu \,}p_{\nu \,}$ . Then the
action is
\[
S=\int_0^Td\tau \,\left( \dot{x}^{\mu \,}\cdot p_{\mu \,}-\frac
12A^{22}G_0^{\mu \nu }\left( x\right) \,p_{\mu \,}p_{\nu \,}\right)
\rightarrow \int_0^Td\tau \frac{G_{\mu \nu }^0\dot{x}^\mu \dot{x}^\nu }{
2A^{22}}
\]
as we have seen above in some examples. From this solution we can construct
a new solution $X^M=F\left( x\right) X_0^M,$ $\,\,P^M=\frac{P_0^M}{F\left(
x\right) }.$ which automatically satisfies the constraints and gives and $
\dot{X}^M\cdot P_M=\dot{x}^{\mu \,}\cdot p_{\mu \,}$. But it also produces a
new metric that differs from the previous one by a conformal factor
\[
S=\int_0^Td\tau \frac{G_{\mu \nu }\dot{x}^\mu \dot{x}^\nu }{2A^{22}},\quad
G_{\mu \nu }=F^2\left( x\right) G_{\mu \nu }^0,\quad any\,\,F\left( x\right)
\]
For example the free particle solution in flat Minkowski space $G_{\mu \nu
}^0=\eta _{\mu \nu }$, gives the conformal particle solution with $G_{\mu
\nu }=F^2\left( x\right) \eta _{\mu \nu }$.

As seen from eq.(\ref{lmnclass}) the classical gauge invariant $L^{MN}$ are
the same in the two models related to each other by a conformal factor.
However, this is not so in the quantum theory in which operators must be
ordered. In \cite{dualsusy} we argued that there is an ordering of the
quantum generators in the new theory that is simply related to the quantum
generators in the old theory, and that the new theory automatically has the
same Casimir eigenvalues as the old theory, in agreement with the general
prediction obtained in the SO$\left( d,2\right) $ covariant quantization.

\section{Acknowledgments}

I thank Franco Iachello for asking the question of whether one can find SO$
\left( d\right) $ invariant Hamiltonians besides the $1/r$ potential. I also
thank the organizers of this conference for their support. This research was
partially supported by the US. Department of Energy under grant number
DE-FG03-84ER40168.


\begin{thebibliography}{99}
\bibitem{ibtokyo} I. Bars, ``Duality and hidden dimensions'', in the
proceedings of the conference {\it Frontiers in Quantum Field Theory},
Toyonaka, Japan, Dec. 1995, Ed. (Itoyama et. al), page. 52, hep-th/9604200;
I. Bars Phys. Rev. {\bf D54 }(1996) 5203.

\bibitem{duff} M. Duff and M.P. Blencowe, Nucl. Phys. {\bf B310} (1988) 387.

\bibitem{ftheory} C. Vafa, Nucl. Phys. {\bf B469} (1996) 403.

\bibitem{stheory} I. Bars, Phys. Rev. {\bf D55} (1997) 2373; I. Bars,
``Algebraic Structures in S-Theory'', hep-th/9608061, lectures in Second
Sakharov conf. 1996, and Strings-96 conf.

\bibitem{14d} I. Bars, Phys. Lett. {\bf B403} (1997) 257.

\bibitem{martinec} D. Kutasov and E. Martinec, Nucl. Phys. {\bf B477} (1996)
652; Nucl. Phys. {\bf B477} (1196) 675; E. Martinec, ``Geometrical
structures of M-theory'', hep-th/9608017.

\bibitem{sezgin} H. Nishino and E. Sezgin, Phys. Lett. {\bf B388} (1996)
569; E. Sezgin, Phys. Lett. {\bf B403 (}1997) 265; H. Nishino,
hep-th/9710141.

\bibitem{sentropy} I . Bars, Phys. Rev. {\bf D55 }(1997) 3633.

\bibitem{ibkounnas} I. Bars and C. Kounnas, Phys. Lett. {\bf B402} (1997)
25; Phys. Rev. {\bf D56} (1997) 3664.

\bibitem{spartstring} I. Bars and C. Deliduman, Phys. Rev. {\bf D56} (1997)
6579.

\bibitem{sparticles} I. Bars and C. Deliduman, ``Gauge principles for
multi-superparticles'', hep-th/9710066, Phys. Lett. {\bf B417} (1998)
240-246.

\bibitem{sezrudy} I. Rudychev and E. Sezgin, ``Superparticles in $D>11$
Revisited'', hep-th/9711128.

\bibitem{dualconf} I. Bars, C. Deliduman and O. Andreev, ``Gauged duality,
Conformal Symmetry and Spacetime with two times'', hep-th/9803188, Phys.
Rev. D58 (1998) 066004.

\bibitem{dualH} I. Bars, ``Conformal Symmetry and Duality between free
particle, H-atom and harmonic oscillator'', hep-th/9804028, Phys. Rev. D58
(1998) 066006.

\bibitem{dualsusy} I. Bars and C. Deliduman, ``Gauge symmetry in phase space
with spin, a basis for conformal symmetry and duality among many
interactions'', hep-th/9806085, to appear in Phys. Rev. D

\bibitem{maldacena} J. Maldacena, hep-th/9711200.

\bibitem{sp2N} I. Bars, ``Sp$\left( 2N\right) $ Duality, Multi-particles and
Multi-times'', to be published.

\bibitem{jarvis} For a recent discussion of BRST methods applied to
particles see, P. Jarvis, A. Bracken, S. Corney and I. Tsohantjis,
``Realizations of particle states via cohomologies: Algebraization of
BRST-BFV covariant quantization'', in Proc. of 5th Wigner Symposium, Vienna,
Austria, 1997, and also these proceedings.

\bibitem{gauges} I. Bars, ``Hidden symmetry and the lifting of
one-time-physics to two-time-physics'', to be published.
\end{thebibliography}
\end{document}